# Ballistic Transport Enhanced Heat Convection at Nanoscale Hotspots

Yanru Xu[1],†, Shen Xu[2],†, Jingchao Zhang[3],†, Jianshu Gao[1], Shugang Deng[1], Wenxiang Liu[1], Xinwei Wang[4], Xin Zhang[5], Yanan Yue[1,6],∗

[1]School of Power and Mechanical Engineering, Wuhan University, Wuhan, Hubei 430072, China.
[2]School of Mechanical and Automotive Engineering, Shanghai University of Engineering Science, Shanghai 201620, China.
[3]NVIDIA AI Technology Center (NVAITC), Santa Clara, CA 95051, USA.
[4]Department of Mechanical Engineering, Iowa State University, Ames, IA 50010, USA.
[5]Department of Mechanical Engineering, Boston University, Boston, MA 02215, USA.
[6]Department of Mechanical and Manufacturing Engineering, Miami University, Oxford, Ohio 45056, USA.

**ABSTRACT:**

Along with device miniaturization, severe heat accumulation at unexpected nanoscale hotspots attracts wide attentions and urges efficient thermal management. Heat convection is one of the important heat dissipating paths but its mechanism at nanoscale hotspots is still unclear. Here shows the first experimental investigation of the convective heat transfer coefficient at size-controllable nanoscale hotspots. A specially designed structure of a single layer graphene supported by gold nanorods (AuNRs) is proposed, in which the AuNRs generate plasmonic heating sources of the order of hundreds of nanometers under laser irradiation and the graphene layer works as a temperature probe in Raman thermometry. The determined convective heat transfer coefficient is found to be about three orders of magnitude higher than that of nature convection, when the simultaneous interfacial heat conduction and radiation are carefully evaluated. Heat convection thus accounts to more than half of the total energy transferred across the graphene/AuNRs interface. Both the plasmonic heating induced nanoscale hotspots and ballistic convection of gas molecules contribute to the enhanced heat convection. This work reveals the importance of heat convection at nanoscale hotspots to the accurate thermal design of miniaturized electronics, and further offers a new way to evaluate the convective heat transfer coefficient at nanoscale hotspots.

---

†These authors contributed equally to this work.
*Corresponding author: Yanan Yue, E-mail: yyue@whu.edu.cn;

**TOC GRAPHIC**

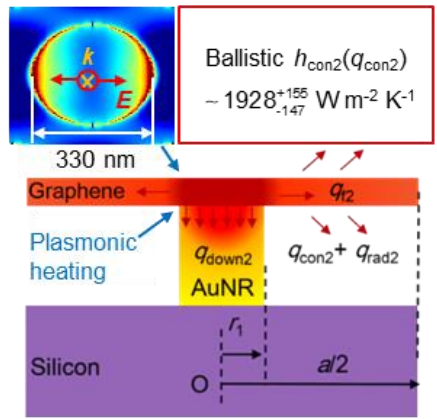

# INTRODUCTION

The increasing power density in condensed integrated circuits and transistors brings an urgent need to explore thermal dissipation mechanism in confined spaces.[1, 2] Besides mass heat accumulation, unexpected hotspots occurring at micro/nanostructures could strongly elevate local temperature and cause material deterioration and even device failure. When the characteristic length of these hotspots are at the scale of the surrounding media's phonon mean free path (MFP), the local thermal transport switches from the diffusion regime to the ballistic regime.[3] Studies have shown the frequent occurrence of ballistic behavior of thermal transport at micro/nanoscale hotspots[4], which is greatly different from the macroscale heat transfer. Taking the ballistic heat conduction as an example, the effective thermal conductivity ($\kappa$) of the local media is lower than its bulk value.[5]

Convection thermal transport plays a critical role in heat removal due to the existence of air surrounding the electronic devices.[6] Studies have shown that the convection thermal transport of water flow at microscale differs from macroscale.[7] However, as the dimension is down to the micro/nanoscale, reported results are scarce due to the measurement difficulties at such extreme scales.[8] Raman thermometry is a well-established approach for small-scale temperature probing, in which temperature of the small scale could be accurately measured by evaluating the Raman shift of the characteristic Raman peaks of the local materials excited by Raman excitation laser. The focal diameter of the Raman excitation laser could be as small as 500 nm. For nano-materials and structures smaller than 500 nm having strong Raman scatterings could further increase the spatial resolution for temperature probing. These material-specific Raman peaks could also be captured using the same approach, such as 135 nm diameter porous silicon membrane and quantum dots.[9, 10] The utilization of high-resolution Raman thermometry could also obtain the temperature at nanoscale hotspots surrounded by air molecules.

In this work, a specially designed tip-patterned substrate under Raman laser irradiation is used to generate hotspots and Raman scatterings. Raman thermometry can be combined to create a tip-enhanced Raman thermometry for hotspots generation and measurements.[11] By using experimental and numerical approaches, a tip-enhanced Raman thermometry is created to directly measure the interface thermal conductance and convective heat transfer coefficient at nanoscale hotspots. A periodic Au nanorods array (AuNRs) is fabricated and utilized as resupinate tips to create hotspots based on their plasmonic effect under laser irradiation.[12, 13] The heat transfer mechanism is then investigated in a single layer graphene supported by the AuNRs to reveal the importance of ballistic transport enhanced heat convection at nanoscale hotspots.

## RESULTS AND DISCUSSION

**Plasmonic Enhancement Effect of Au Nanorods Array.**

Preparation of gold nanorods array is detailed in Methods. The array of AuNRs (Au330/Si) with an average diameter of 330 nm, a height of 200 nm, and a periodicity of 450 nm is shown in Figure 1a. A single layer of CVD graphene is transferred onto the top of Au330/Si. Compared with the close contact between graphene and gold sputter-coating film/silicon substrate (AuF/Si), the contact areas between graphene and the patterned substrate are restricted to the tip areas of the gold nanorods. The plasmonic enhancement effect of Au330 is revealed from the Raman intensity difference between graphene on Au330/Si and on AuF/Si (Figure 1f).

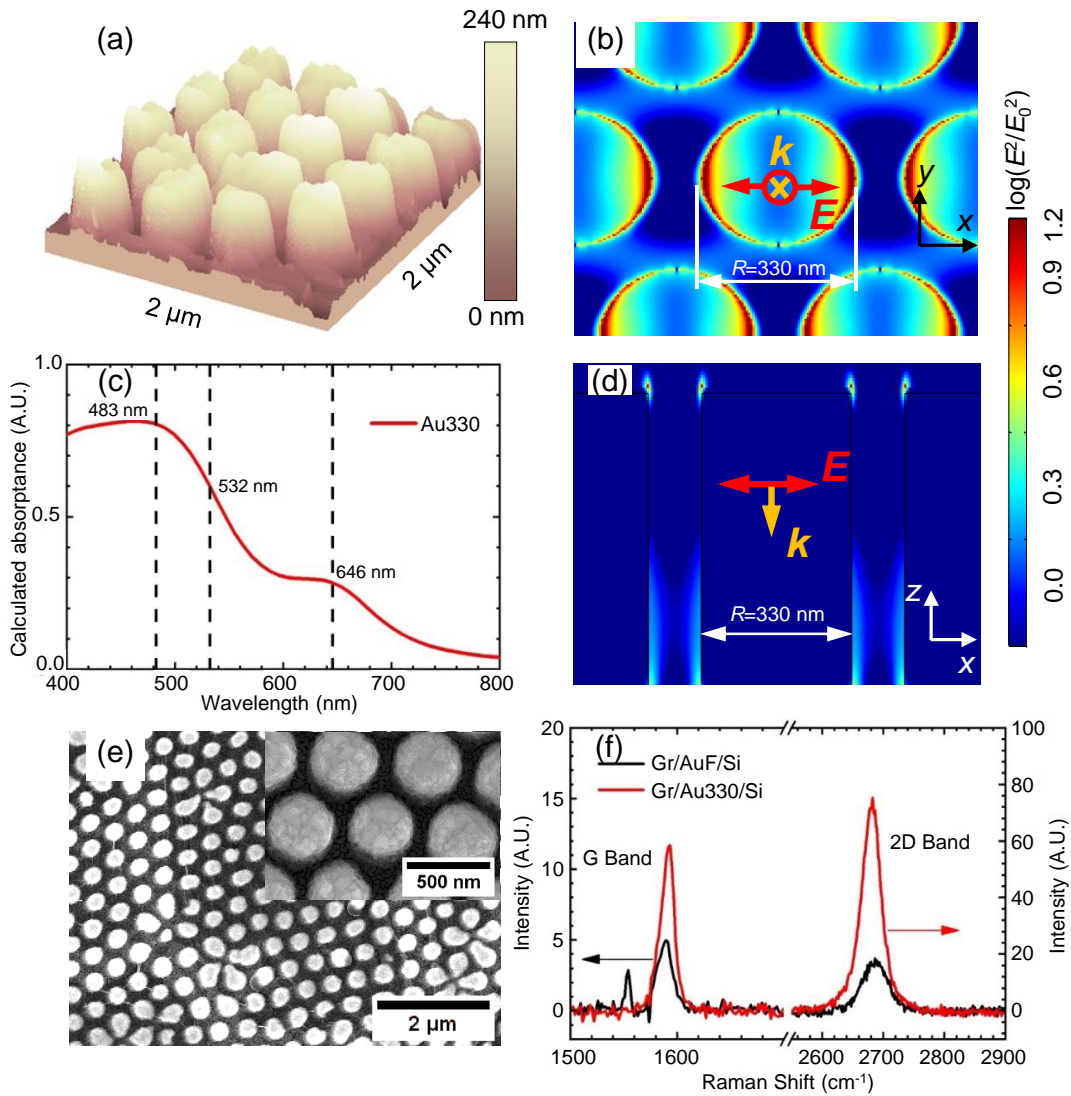

**Figure 1.** Surface enhancement on the Au nanorods. (a) 3D AFM image of Au330/Si showing the height of the Au nanorods. (c) The simulated extinction spectrum of Au330. The electric intensity enhancement distribution (on a log scale) on (b) top surface of gold nanorods and (d) central *xz* plane at 532 nm laser excitation. (e) SEM images of Gr/Au330/Si. (f) Raman spectra of graphene on AuF/Si and Au330/Si. An enhancement factor of 11.8 is observed for G peak of graphene on Au330/Si.

LSPR (localized surface plasmon resonance) generated by Au330 enhances the Raman intensity by a factor of 11.8 and 20.4 for the *G-band* and the *2D-band*, respectively. The enhancement distribution induced by Au330 can be pictured by the simulated electromagnetic field (see Methods for details). The Au330 absorptance (Figure 1c) decreases from 0.8 to 0.04 as the incident wavelengths increasing from 400 to 800 nm with two plasmonic bands. Figure 1b and 1d show that the top surface electric field at the nanorods edge is greatly intensified generating theoretical Raman enhancement factor of 9.3 for *G-band* at 532 nm.[14, 15]

**Interfacial Thermal Conductance Between Au and Graphene.**

The incident laser serves as the exciting and heating source simultaneously in tip-enhanced Raman thermometry. The generated thermal energy in the graphene layer will dissipate to the surroundings through heat conduction, convection and radiation. Schematics of the experimental setup and measuring mechanism for the interfacial thermal conductance between graphene and AuF/Si, Au330/Si are shown in Figure 2a. For Gr/AuF/Si in Figure 2b, the temperature is uniform within the irradiated spot due to the larger focal spot size than the mean free path of phonons in graphene.[16] The main thermal pathway for the generated thermal energy is across the Gr/Au interface with some dissipation through thermal radiation and convection from the top surface of the graphene layer. For Gr/AuNRs/Si, the thermal transport path is slightly different since the graphene layer does not contact closely with the substrate in all areas. Part of graphene is suspended and surrounded by air at both top and bottom surfaces shown in Figure 1e.

Thus, aside from conduction, the generated thermal energies will convect or radiate to surroundings from graphene surfaces at both top and bottom positions. To investigate the convective thermal transport from graphene on Au330/Si, the thermal conduction across the Gr/Au interface must be explored first.

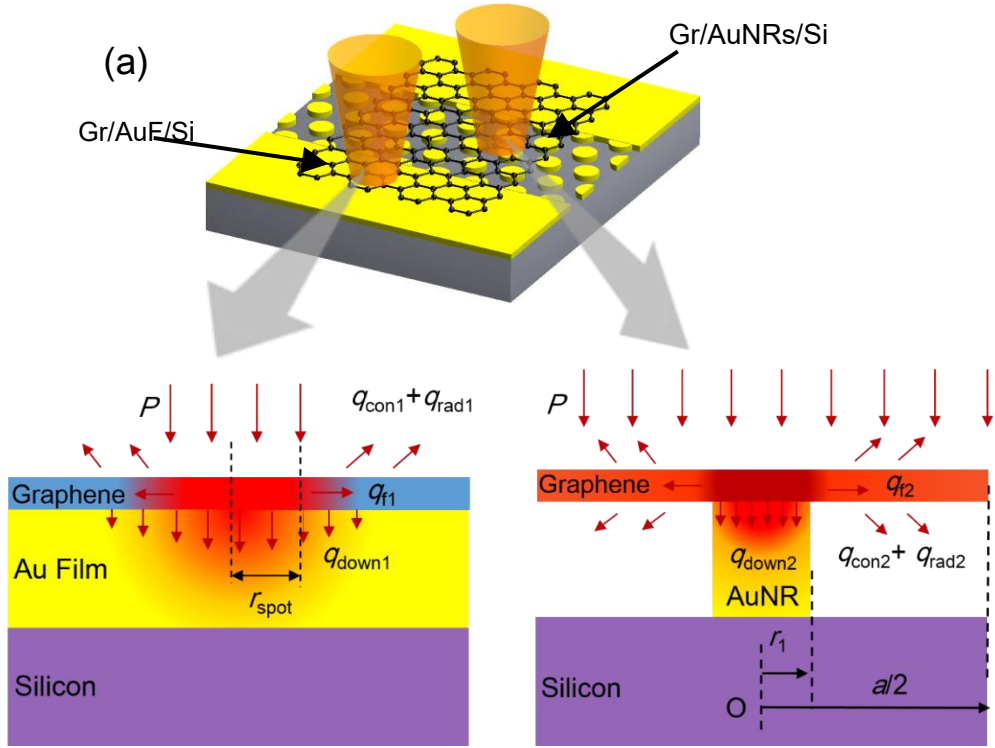

**Figure 2.** Different heat transfer mechanisms in two structures. (a) The schematic of laser heating on Gr/AuF/Si and Gr/AuNRs/Si. Two different mechanisms of thermal transport in (b) Gr/AuF/Si and (c) Gr/AuNRs/Si under the laser heating.

As illustrated in Figure 2a, the abovementioned two samples, Gr/AuF/Si and Gr/AuNRs/Si, are prepared in the same batch, so the roughness of the surface for AuF and AuNRs should be similar. Furthermore, the AuNRs have the same height as AuF and a quiet high density on the Si substrate, the transferred graphene will have a good contact with the AuNRs as it does on the Au film. So we assume that the contact between graphene and the AuNRs is the same as that between graphene and AuF. We first measure the interfacial thermal conductance (Gr/Au) between graphene and the Au film in the Gr/AuF/Si sample. The temperature of graphene on AuF/Si is measured using the steady-state opto-thermal Raman method (see Methods for details). The contour map of Raman intensity of *G* peak against the incident laser power is shown in Figure 3a. To alleviate measurement uncertainties, we measure the linear relationship between peak position in the *G-band* and excitation laser power, rather than directly measuring the temperature at a certain incident power. The fitted slope of -0.029 ± 0.011 cm$^{-1}$ mW$^{-1}$ in Figure 3c together with the calibrated temperature coefficient of -0.021 cm$^{-1}$ K$^{-1}$ for the *G-band*, determines that the average local temperature rise in the laser irradiated region is 41.4 ± 9.5 K at a laser power of 30 mW. The experiment-based simulation is performed to calculate the interface thermal conductance. A well-defined heat transport model for Gr/AuF/Si is developed using a stable numerical finite element (FE) method for the temperature rise of graphene on AuF/Si under laser irradiation at 30 mW (see Methods for details). To best fitted the experimentally measured temperature rise of 41.4 ± 9.5 K of the graphene layer, the interface thermal conductance between graphene and gold $G_{Gr/Au}$ is determined as $1.35^{+0.37}_{-0.33} \times 10^4$ W m$^{-2}$ K$^{-1}$.

**Convective Heat Transfer Coefficient in the Transition Regime over the Nanoscale Hotspots.**

The surface temperature of graphene on Au330/Si is measured using the same Raman method and its Raman intensity contour map against incident power is shown in Figure 3b. Combined with the temperature coefficient of *G-band* Raman shift (-0.021 cm$^{-1}$ K$^{-1}$), the temperature rise of graphene is determined by the slope of the fit of the *G-band* peak position to the laser power (-0.074 cm$^{-1}$ mW$^{-1}$), as shown in Figure 3d. The average local temperature rise within the laser irradiated region is 105.7 ± 2.4 K at a laser power of 30 mW, which is much higher than 41.4 ± 9.5 K for graphene on AuF/Si, indicating that graphene is additionally heated by the Au330-induced plasmonic heating (nanoscale hotspots). Although the near-field effect will enhance the amount of photon absorption and heat generated in the AuNRs, the heat will dissipate rapidly through the silicon substrate due to the high thermal conductivity of both gold and silicon. For graphene on Au330/Si, the contact area of the graphene/Au330 interface is reduced due to the discontinuity of the upper surface of the Au330 layer, and the generated heat can only be conducted through the contact area. The low thermal conductance at the loosely contacted interface limits the heat dissipation and raises the temperature of the graphene layer. By improving the photon absorption in graphene, the confined electric field in the local region around the nanorods provides an opportunity to generate nanoscale hotspots in graphene.

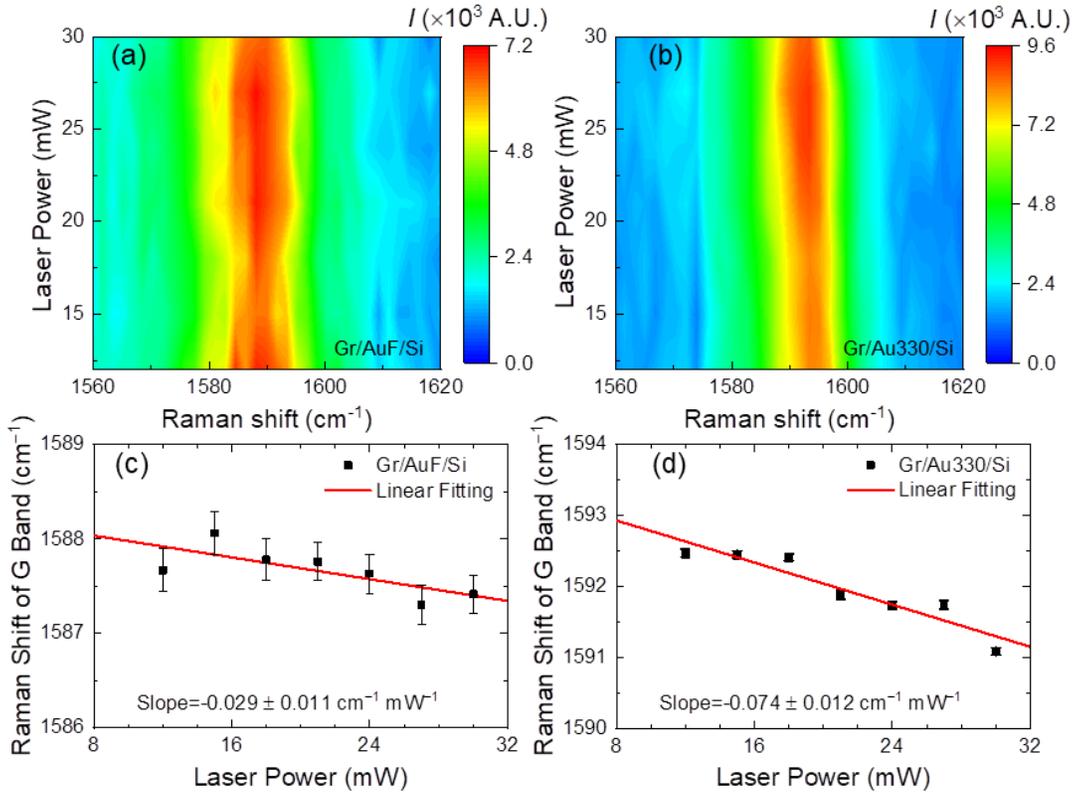

**Figure 3.** Raman based temperature measurement. The contour map of Raman peak for G-band vs. laser power for (a) Gr/AuF/Si and (b) Gr/Au330/Si. The inset depicts a G-band line from laser power of 12 mW to 30 mW.

The physical model in FE simulation of heat transport in graphene is shown in Figure 2c (see Methods for details). The interfacial heat transfer between graphene and the AuNRs is set to be $1.35^{+0.37}_{-0.33} \times 10^4$ W m$^{-2}$ K$^{-1}$, referring to the one between graphene and the Au film, to evaluate the thermal convection coefficient. Convective boundary conditions are applied to the exposed surfaces at the top and bottom of the graphene. In comparison, the radiative heat loss is negligible, as it is estimated to be only 0.002-0.003% of the total heat dissipation of Gr/Au330/Si. When the average temperature rise of graphene in the laser irradiated region is similar to the measured temperature rise (105.7 ± 2.4 K), the effective average convective heat transfer coefficient (*h*) in the irradiated region is 1928 $^{+155}_{-147}$ W m$^{-2}$ K$^{-1}$. It is particularly higher than the bulk value (10 W m$^{-2}$ K$^{-1}$).

**Discussion.**

The hotspots generated by the LSPR effect correspond to the size of the nanorods. In tip-enhanced Raman thermometry, they work as nanoheaters that dissipate heat into the air molecules surrounding the nanorods. When the size of these nanoheaters is close to the mean free path of air molecules at room temperature and pressure, a change in their size will alter the convective heat transfer coefficient *h* by several orders of magnitude.[7] To confirm the size dependence of *h*, we fabricated gold nanorods of another size on a silicon substrate with a diameter of 240 nm, a height of 100 nm (hereafter Au240/Si), and a period of 450 nm, in accordance with Au330 (see Supplementary Information for details). At a laser power of 30 mW, the temperature rise of graphene on Au240/Si reaches 81.4 ± 3.8 K. The average *h* over the entire laser irradiated region is then determined to be 1793$^{+157}_{-159}$ W m$^{-2}$ K$^{-1}$. Similar to the average *h* on the top surface of Gr/Au330/Si, it is about three orders of magnitude higher than nature convection.

The convective heat transfer coefficient of graphene on Au330 is slightly higher than that of graphene on Au240 because the total heated area of graphene due to the LSPR effect is different for Au240 and Au330. They are lower than the theoretical predictions reported between the nanomaterial and the gaseous environment. The LSPR effect of graphene is not considered, which may lead to an underestimation of graphene heating. In addition, it may be related to the geometric differences between 2D and 1D materials.[17, 18]

The combination of nanoscale hotspot heating and ballistic convection effect of gas molecules may be responsible for the high values of the measured convective heat transfer coefficients. The LSPR-induced nanoscale hotspot puts air convection in transition regime and transfers heat in a ballistic manner. The heat transfer government of the surrounding air molecules over the nanostructure-induced hotspots can be defined by *Kn* number, which is the ratio of the mean free path $\lambda$ of air molecules[19] to the characteristic length of the hotspot. In the case of open-air measurements of Gr/Au330/Si, $\lambda$ of air molecules is 80 nm and the *l* of gold nanorods is 330 nm. The resultant *Kn* is 0.24, which lies in the range of 0.01 to 10, indicating that the airflow regime over the hotspot belongs to the transition regime. Since the size of the hotspot is similar to the mean free path of air molecules, the probability of ballistic heat conduction of free air molecules may increase. Heat conduction through air molecules plays an important role instead of advection-based heat transfer[18], leading to enhanced thermal convection. The evaluation of heat dissipation based on the determined heat transfer coefficients demonstrates that convective heat loss accounts for 53%-100% of the total heat conduction from graphene to Au330/Si and Au240/Si through the interface. This high percentage indicates that the convective heat loss from the nanoscale hotspots in the graphene layers to the surrounding air is significant and cannot be neglected in the precise design of graphene-based thermal management systems. Meanwhile, the tip-enhanced Raman thermometry developed in this work offers a new methodology for measuring the convective heat transfer coefficient of the nanoscale hotspots.

## CONCLUSION

In summary, we measured the convective heat transfer coefficients of air molecules in the transition regime over the nanoscale hotspot based on our self-developed tip-enhanced Raman thermometry. The experimental results show that the gold nanorod-induced near-field effect (LSPR) will enhance the Raman intensity and incident energy absorption of the graphene on top of AuNRs, resulting in an LSPR-induced hotspot at the graphene layer. The temperature increase in the graphene layer has been experimentally measured and used to reconstruct the heat transfer model in the sample and to determine the convective heat transfer coefficient over the hotspots. The determined average convective heat transfer coefficients, $1928^{+155}_{-147}$ W m$^{-2}$ K$^{-1}$ for Gr/Au330 and $1793^{+157}_{-159}$ W m$^{-2}$ K$^{-1}$ for Gr/Au240, indicate convective heat transfer in the transition system. Our self-developed tip-enhanced Raman thermometry provides a new method to experimentally quantify ballistic heat transport from nanoscale hotspots. The measurement results can be used for better thermal design and management at the micro/nanoscale.

# EXPERIMENTAL METHODS

## Preparation of Au Nanorods Array

Template-assisted lithography is used to prepare the gold nanopattern on the silicon substrate. The commercial ultrathin alumina membrane (UTAM) with a thickness of 650 nm, diameter of 350 nm, and periodicity of 450 nm was used as a template. The polymethyl methacrylate (PMMA) was used as support.[20] A single-layer graphene (Gr) with the dimension of 1 × 1 cm$^2$ was transferred onto the prepared Au330/Si. The uncovered silicon substrate was simultaneously deposited with continuous Au film (AuF/Si). Sample preparations are described in Figure S1. The prepared AuNRs on silicon (Au330/Si) have an average diameter of 330 nm and a periodicity of 450 nm, as shown in Figure S2.

## Electromagnetic Field Simulation of Au Nanorods Array

The extinction curve of Au330 is calculated using finite-difference time-domain (FDTD) simulations. Details of the simulation model are shown in the Supplementary Information. To determine the absorption of incident energy (extinction, $A$) by the gold layer, we simulated reflection ($R$) and transmission ($T$) on the gold film. We calculate the extinction by $A = 1 - R - T$. The electric field intensity distribution around Au330 is also simulated by using the FDTD model described above. According to the electromagnetic enhancement mechanism of surface-enhanced Raman scattering, the theoretical Raman enhancement factor of Au330-induced graphene is expressed as the ratio of the local electrical intensity to the incident electrical intensity ($E^2/E_0^2$), which is averaged over the top surface of the nanorods.

## Measurements of Temperature Rise of Graphene

Raman thermometry was used for temperature measurement of graphene. A Raman spectrometer (B&W Tek) with a 532 nm diode laser is used for temperature coefficient calibration of graphene Raman signal. The focusing laser spot has a diameter of ~50 μm at the tested surface with a power of 30 mW. Accurate temperature control was achieved by a ceramic heater within the range of 300 to 365 K. An integration time of 240 seconds was used to ensure the temperature measurement accuracy. Each experiment was conducted twice, with each spectrum collected three times for averaging. The Lorentzian function is used to fit the peak position of the *G-band* to determine the graphene temperature (Figure S4). The peak frequency of *G-band* has a red shift with increasing temperature (Figure S5). The temperature coefficient of *G*-peak of our sample is determined as -0.021 cm$^{-1}$ K$^{-1}$.[21] To alleviate the measurement uncertainties, the change in Raman shift against laser power was used to extract the graphene temperature rise. For temperature rise measurement of graphene on AuF/Si, the laser power per unit area was adjusted from 0.006 to 0.015 mW μm$^{-2}$. The integration time was varied from 400 to 160 seconds. For the temperature measurement of graphene on AuNRs, selections of the monitored positions are shown in Supplementary Information. The correlations of Raman shift with laser power were used to extract the temperature rise. The laser power was adjusted from 12 to 30 mW. The corresponding power intensities ranged from 0.006 to 0.015 mW μm$^{-2}$. The integration time was varied from 50 to 20 s to obtain strong Raman signals. Each experiment was conducted twice, with each spectrum collected three times for averaging.

**The Experiment-based Simulation of Graphene**

A schematic of the finite element model between graphene and the continuous gold film is shown in Figure S6(a). Simulation details are provided in the Supplementary Information. In the heat transport model for Gr/AuF/Si, thermal convection boundaries were applied to the top surface of graphene with a thermal convection coefficient of 10 W m$^{-2}$ K$^{-1}$. The correlation between the interfacial thermal conductance $G_{Gr/Au}$, and the surface temperature in the simulation model determines the real actual value of $G_{Gr/Au}$. When the surface temperature of graphene matches the experimental results, $G_{Gr/Au}$ of $1.35_{-0.33}^{+0.37} \times 10^4$ W m$^{-2}$ K$^{-1}$ is obtained. This value is the same as that of the unconstrained graphene/4-H SiC interface[22] but is much lower than that reported for the intercalated graphene interface.[17] It indicates the existence of a gap between graphene and AuF/Si. Molecular dynamics simulations also confirm that the conductance decreases exponentially with increasing gap thickness. This separation weakens the interatomic forces and energy coupling between the two materials. Ripples and wrinkles in graphene as well as chemical residues introduced during the preparation process will be the main cause of interfacial separation. In addition, the different thermal expansion behavior of graphene and gold may cause thermal mismatch[22] at the interface and further increase their separation distance. Schematic diagram of the simulated model of graphene on Au330/Si is depicted in the Supplementary Information. The convective heat transfer coefficient ($h$) around graphene on gold nanopatterns was determined when the simulated average temperature rise of graphene in the laser irradiated region matched the experiment ($105.7 \pm 2.4$ K).

## Supporting Information

The Supporting Information is available free of charge at: advances_supplementary_materials_.docx

## Corresponding Authors

**Yanan Yue -** *School of Power and Mechanical Engineering, Wuhan University, Wuhan, Hubei 430072, China;* Email: yyue@whu.edu.cn

## Notes

The authors declare no competing financial interest.

## ACKNOWLEDGEMENTS

The authors acknowledge the financial support from the National Key Research and Development Program (No.2019YFE0119900), National Natural Science Foundation of China (No. 52076156 & No. 52106220) and Fundamental Research Funds for the Central Universities (No. 2042020kf0194).